\documentclass[conference]{IEEEtran}
\usepackage[absolute]{textpos}
\IEEEoverridecommandlockouts
\usepackage{cite}
\usepackage{amsmath,amssymb,amsfonts}
\usepackage{graphicx}
\usepackage{textcomp}
\usepackage{xcolor}

\usepackage{mathptmx}
\usepackage{array}
\usepackage{tabularx}
\usepackage{algorithm}
\usepackage{algpseudocode}
\usepackage{booktabs}
\usepackage{url}
\usepackage{makecell}
\usepackage{multirow}
\usepackage{multicol}
\usepackage[bookmarks=false]{hyperref}

\def\BibTeX{{\rm B\kern-.05em{\sc i\kern-.025em b}\kern-.08em
    T\kern-.1667em\lower.7ex\hbox{E}\kern-.125emX}}
\begin{document}
    
\begin{textblock*}{18cm}(1.5cm,0.5cm)
\noindent\centering\footnotesize
\copyright~2026 IEEE. Personal use of this material is permitted. Permission from IEEE must be obtained for all other uses, in any current or future media, including reprinting/republishing this material for advertising or promotional purposes, creating new collective works, for resale or redistribution to servers or lists, or reuse of any copyrighted component of this work in other works. \\
\textbf{The official Version of Record is available at IEEE Xplore (DOI: \href{https://ieeexplore.ieee.org/}{10.1109/ICKG70389.2026.00009} resolving soon).}
\end{textblock*}

\title{SuSi-RAG: A Hybrid Graph RAG for Equipment Maintenance \\
\thanks{This study was partially supported by the EU Project SMARTY (GA 101140087). $\dagger$ The authors contributed equally to this work.}
}

\author{\IEEEauthorblockN{\hspace{0ex}Hansi Zhang$\dagger$}
\IEEEauthorblockA{\hspace{0ex}\emph{Robert Bosch GmbH}\\
\hspace{0ex}Shanghai, China\\
\hspace{0ex}hans.zhang@cn.bosch.com}
\and
\IEEEauthorblockN{\hspace{0ex}Wilma Johanna Schmidt$\dagger$}
\IEEEauthorblockA{\hspace{0ex}\emph{Robert Bosch GmbH}\\
\hspace{0ex}Renningen, Germany\\
\hspace{0ex}\emph{Freie Universität Berlin}\\
\hspace{0ex}Berlin, Germany\\
\hspace{0ex}wilma.schmidt@bosch.com}
\and
\IEEEauthorblockN{\hspace{0ex}Wanlin Yang\hspace{8ex}}
\IEEEauthorblockA{\hspace{0ex}\emph{Robert Bosch GmbH}\hspace{8ex}\\
\hspace{0ex}Shanghai, China\hspace{8ex}\\
\hspace{0ex}\emph{Fudan University}\hspace{8ex}\\
\hspace{0ex}Shanghai, China\hspace{8ex}\\
\hspace{-6ex}wanlin.yang@cn.bosch.com}\hspace{8ex}
\and
\IEEEauthorblockN{\hspace{15.5ex}Xiaozhi Shen\hspace{8ex}}
\IEEEauthorblockA{\emph{\hspace{15.5ex}Robert Bosch GmbH}\hspace{8ex}\\
\hspace{15.5ex}Suzhou, China\hspace{8ex}\\
\hspace{9.5ex}xiaozhi.shen@cn.bosch.com}\hspace{8ex}
\and
\IEEEauthorblockN{\hspace{4ex}Adrian Paschke}
\IEEEauthorblockA{\hspace{4ex}\emph{Fraunhofer Institute FOKUS}\\
\hspace{4ex}Berlin, Germany \\
\hspace{4ex}\emph{Freie Universität Berlin}\\
\hspace{4ex}Berlin, Germany \\
\hspace{4ex}adrian.paschke@fokus.fraunhofer.de}
\and
\IEEEauthorblockN{\hspace{5ex}Evgeny Kharlamov\hspace{5ex}}
\IEEEauthorblockA{\hspace{5ex}\emph{Robert Bosch GmbH}\hspace{5ex}\\
\hspace{5ex}Renningen, Germany\hspace{5ex}\\
\hspace{5ex}\emph{University of Oslo}\hspace{5ex}\\
\hspace{5ex}Oslo, Norway\hspace{5ex}\\
\hspace{5ex}evgeny.kharlamov@de.bosch.com}\hspace{5ex}
}

\maketitle

\begin{abstract}
In the manufacturing domain, equipment maintenance tasks depend on expert knowledge. Their documentation is often stored in heterogeneous formats and sources. The enhancement of \emph{LLMs} with \emph{Retrieval-Augmented Generation (RAG)} is promising for bridging the scattered knowledge, but traditional vector-based RAG shows weaknesses with two characteristics which are common in real maintenance queries: complex causal chains and cross-domain synthesis. To address this, we propose \emph{SuSi-RAG}, a hybrid Graph RAG solution, that matches a user query and each maintenance record using subgraph similarity. 
This work shows an \emph{in-use} system: \emph{SuSi-RAG} has run on an active Bosch production line for more than \emph{eight} months, handling \emph{10,664} question-answer interactions and demonstrates upper bound answer accuracy of \emph{96.25\%} (confirmed error rate 3.75\%). In a proof-of-concept, \emph{SuSi-RAG} improves top-\emph{1} retrieval accuracy from 61.5\% to 96.2\% and expert-rated answer quality from 44.9\% to 84.1\% against a vector-based baseline. We present an integrated operational feedback loop to update \emph{SuSi-RAG}.
\end{abstract}


\section{Introduction}\label{sec_introduction}
\begin{figure}[htbp]
\centerline{\includegraphics[width=0.49\textwidth]{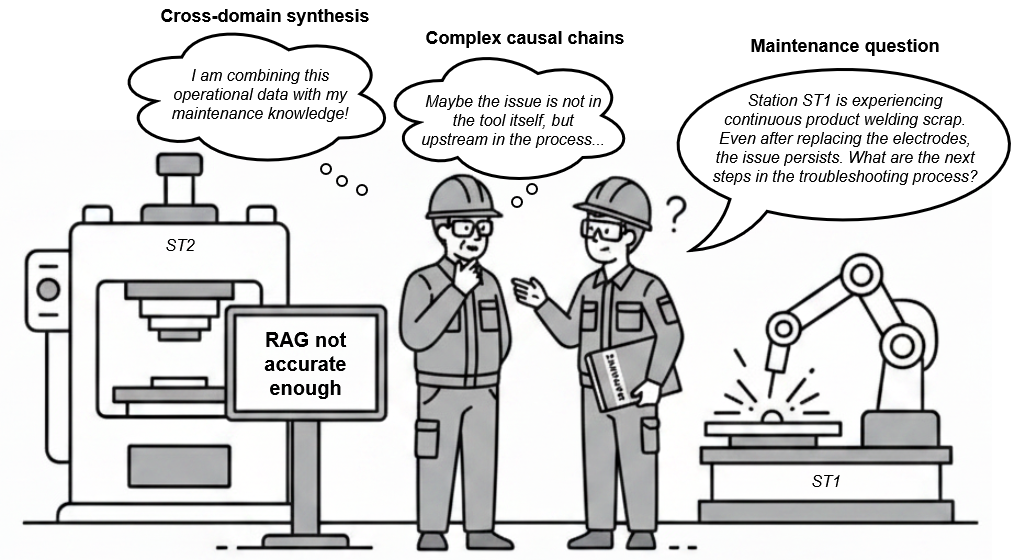}}
\caption{Motivation of this work: accessing relevant knowledge for equipment maintenance is crucial for solving the incident, yet it is tedious and non-trivial.}
\label{fig:motivation}
\vspace{-1.5ex}
\end{figure}

More than a decade of Industry 4.0~\cite{lasi2014industry} brought benefits for manufacturing~\cite{lin2023industry}, but also increasingly complex shopfloor equipment and industrial machinery with interconnected soft- and hardware. 
Efficient and high-quality maintenance of this equipment is economically critical for manufacturing companies~\cite{DBLP:journals/cii/SilvestriFISC20}, yet, the required expertise is scarce as experienced engineers possess valuable expert knowledge, while inexperienced personnel needs support in difficult tasks. 
One exemplary maintenance task is triggered by this question: 
\begin{quote}
\emph{Station ST1 is experiencing continuous product welding scrap. Even after replacing the electrodes, the issue persists. What are the next steps in the troubleshooting process?}
\end{quote}
Keyword-based retrieval often fails here as it might retrieve documents related to \emph{electrode maintenance}, despite that this step was already unsuccessfully attempted. 
A sophisticated approach is needed to follow a causal chain, e.g., \emph{welding scrap} to \emph{position offset} to \emph{gripper status} to \emph{cylinder failure}, and harmonize information across mechanical, process and control domains. \textbf{Complex causal chains (C1)} and \textbf{cross-domain synthesis (C2)} are two common challenges at large manufacturing companies such as Bosch. We illustrate the challenges in Figure~\ref{fig:motivation}.
Semantic technologies, such as ontologies and knowledge graphs (\emph{KGs)}, have proven effective across industrial applications~\cite{iglesias2022ontology,yahya2024benchmark,hofer2024construction,kalayci2020semantic,DBLP:conf/eurocast/ChhetriAFGGG22}. Graph structures enable multi-hop search for causal chains (\textbf{C1}) and maintenance knowledge can be parsed to graph structure with the help of ontologies and \emph{KGs}. 
Yet, accessing this knowledge in practice is tedious and non-trivial as no sufficiently accurate solution exists for our scenario.
A promising direction are hybrid architectures of graph- and LLM-based components (e.g., \cite{hu2025grag}, \cite{xu2024retrieval}).
Hence, we ask the following research question: \\
\noindent
\emph{How can we combine operational data and domain knowledge in a RAG-based solution to answer real maintenance queries and does such a solution prove helpful in production use?} \\
Unlike existing \emph{KG} integration approaches for \emph{RAG} systems, we propose with \emph{SuSi-RAG} a domain-tailored retrieval method to this question. While user queries and maintenance documents are represented as maintenance graphs, a similarity score with expert-weighted entity types and a proportionality harmonization of incomplete queries selects the most relevant records. A balancing step with two common methods ensures robustness and selects the final output candidate. 
In contrast to most similar studies, the evaluation of our industry- and application-oriented work is two-fold: a controlled \emph{proof-of-concept (PoC)} and an \emph{in-use} analysis on a real production line.
Our contributions are:
\begin{itemize}
    \item \textbf{A subgraph-similarity retrieval mechanism} for maintenance events with expert-weighted entities, a harmonization of only partial queries, separated pattern and textual handling to improve search relevance of retrieved items.
    \item \textbf{An in-use analysis} of the implemented, deployed solution for a timeframe of \emph{eight} months on a selected active Bosch production line with \emph{10,664} interactions and expert-confirmed error rate. In addition, a feedback loop mechanism is presented.
    \item \textbf{Insights from Practice} on parameter configurations for reproducibility, and lessons-learned from the deployment to obtain reliable shopfloor results.
    \item \textbf{A quantitative comparison of \emph{SuSi-RAG}} against a vector-based RAG with favourable results: \emph{SuSi-RAG} nearly doubles the performance, increasing retrieval accuracy from 61.5\% to 96.2\% and expert-rated answer quality from 44.9\% to 84.1\%.
\end{itemize}
The remainder of the paper is organized as follows. 
Next, we discuss related research and identify the research gap on our scenario in Section~\ref{sec_relatedwork}. 
In Section~\ref{sec_usecase}, we introduce the \emph{Intelligent Equipment Maintenance Use Case} at Bosch and explain the two main challenges with exemplary maintenance questions. 
In Section~\ref{sec_solution}, we clarify why a hybrid Graph RAG design is needed, and subsequently, describe ontology, \emph{KG} and \emph{SuSi-RAG} with each module in detail. 
In Section~\ref{sec:eval}, we provide configuration settings and evaluate \emph{SuSi-RAG} in a two-fold manner, with a quantitative \emph{PoC} analysis and a continuous in-use analysis of the solution to address. We close this section by discussing the limitations of our approach.
In Section~\ref{sec_lessonslearned}, we present our lessons learned, and, conclude our work in Section~\ref{sec_conclusion} with an outlook on future steps.

\vspace{-1.5ex}
\section{Related Work}\label{sec_relatedwork}
Next, we discuss current research on relevant \emph{RAG} variants as well as \emph{KG} adoption in the manufacturing domain. \\
\noindent
\textbf{Vector-based RAG.} 
A main strategy to augment LLMs is Retrieval-Augmented Generation (RAG)~\cite{gao2023retrieval}, which integrates retrieval-based approaches with generative models (e.g., LLMs)~\cite{lewis2020retrieval} and can improve the factuality, transparency, updateability of the LLM. Domain-specific data and data updates can be integrated in the retrieval component. However, vector-based retrieval has limited contextual awareness~\cite{peng2025graph}, which is insufficient for complex causal chains (\textbf{C1}). \\
\noindent
\textbf{Graph RAG.} 
To overcome this, graph RAG~(\cite{edge2024local,peng2025graph,xu2024retrieval}) leverages the structured nature of graphs. 
First works explore RAG with subgraph retrieval, e.g., \emph{SubgraphRAG}, adjusting the graph size to ``the query's needs and the downstream LLM's capabilities''~\cite{li2024simple}. In another work, a \emph{divide-and-conquer} strategy searches for the optimal subgraph~\cite{hu2025grag}. Our approach differs with the expert-assigned entity weights for the fixed subgraph schema, separate pattern- and text-based entity value handling, and a penalty factor to balance the similarity score of queries with few keywords. \emph{SuSi-RAG} harmonizes the subgraph-similarity ranking with two more methods to increase robustness of the approach, see Section~\ref{sec_solution}. \\
\noindent
\textbf{Hybrid RAG.} Hybrid RAG frameworks combine RAG and graph RAG approaches to benefit from both architectures~\cite{sarmah2024hybridrag}. Xu \emph{et al.}~\cite{10.1007/978-981-97-5678-0_34} compare strategies on how to incorporate KGs for knowledge retrieval. Motivated by challenges of long-text inputs, the authors propose a KG-based dual-pathway approach to ``refine retrieval quality and yield more accurate and coherent outputs''~\cite{10.1007/978-981-97-5678-0_34}. 
Their approach ameliorates the prompts by ``explicit comparison between the KG and the extracted entity relations from queries''~\cite{10.1007/978-981-97-5678-0_34}. The authors mention great challenges for cross-domain queries as well as query performance for large-scale data. We explore especially the cross-domain challenge with our work. \\
\noindent
\textbf{Industrial KG and RAG Setups.} \emph{KGs} are proven to integrate manufacturing data~\cite{yahya2024benchmark,kalayci2020semantic}. Recently, also \emph{RAG} setups in manufacturing gained more attention~\cite{zhou2026lightweight,CHANDRASEKHAR2024100232}.

In summary, to the best of our knowledge, no prior work proposes a hybrid graph RAG combining an expert-weighted maintenance event subgraph similarity matching algorithm, enhanced with a vector store, and evaluates this in-use. Next to specific details of weighted entities and a penalty factor for queries with only partial matchings, \emph{SuSi-RAG} evaluates multiple relevant features under real production conditions.

\section{Intelligent Equipment Maintenance Use Case and Requirements}\label{sec_usecase}
We present use case and requirements of the target solution.

\subsection{Use Case Description}
\emph{Maintenance} describes ``the function that monitors and keeps plant, equipment, and facilities working''~\cite{manzini2010maintenance}. Focus of this work is \emph{equipment maintenance}. Maintenance personnel analyzes operational data to identify potential causes and perform effective maintenance.
A notable challenge is the knowledge gap between an experienced engineer relying on valuable, extensive maintenance experience and a junior colleague who might struggle in root cause identification and troubleshooting.
To bridge this gap, maintenance personnel follows standard procedures, documented and enriched with non-trivial insights from senior engineers. 
While maintenance documents can provide much guidance for junior staff, they are also the basis for continuous improvement of maintenance routines. 
A previously deployed \emph{RAG} solution at Bosch~\cite{zhang2025knowledge} shows accuracy limitations which we quantify next, based on several test runs: the average retrieval accuracy is limited to \emph{61.5\%}, which hinders the solution’s ability to identify the most relevant source documents. The generation accuracy is even lower, at \emph{44.9\%}, leading to incomplete or incorrect answers to queries.
Table~\ref{tab:typical_questions} shows two typical equipment maintenance questions with expert gold answers. Question~1 illustrates multi-hop causal-chain queries with complex relations and entities (\textbf{C1}). Multi-hop queries are queries that require linking fragmented knowledge across heterogeneous documents. These type of queries are typical in equipment maintenance tasks, but also in other areas such as healthcare (see e.g.,~\cite{wang2025llm}) or finance (see e.g.,~\cite{karim2024scalable}). Question~2 is an example for cross-domain synthesis, linking operational data with domain knowledge from maintenance procedure guidelines and potential incident causes, and further synthesizing elements from the mechanical domain with process parameters (\textbf{C2}).
\begin{table}[t]
\caption{Two typical equipment maintenance questions. \emph{Key: * answers are provided by a stakeholder group of maintenance experts; Ch.: Challenge, C1: Complex Causal Chains; C2: Cross-Domain Synthesis.}}\label{tab:typical_questions}
\centering
\renewcommand{\arraystretch}{1.2}
\begin{tabular}{p{0.07\columnwidth} p{0.36\columnwidth} p{0.42\columnwidth}}
\toprule
\textbf{Ch.} & \textbf{Question (shortened)} & \textbf{Gold Answer (shortened)}* \\
\midrule
\textbf{C1} & Station \emph{ST1} is experiencing continuous product welding scrap. Even after replacing the electrodes, the issue persists. What next? & The next troubleshooting steps are: 1.~Position Observation: Observe physical offset in welding position. 2.~ Inspect shuttle position and motor pin correction cylinder. 3.~Root Cause Localization at grippers \emph{G1}, \emph{G2}. If they fail to tighten, replace. 4.~After replacement, monitor sensor signal status. \\
\hline
\textbf{C2} & Station \emph{ST2} reports continuous test failures; press cannot reset (timeout \emph{T1}, error \emph{E1}). How to handle? & 1.~Hardware check of power pin status. 2.~If press fails to reach home position, move back with debugging software \emph{ABC}. 3.~Check and modify press limit settings within process control parameters; compare similar stations (e.g., \emph{ST3}) to optimize to a more stable range.  \\
\bottomrule
\end{tabular}
\vspace{-4ex}
\end{table}

\subsection{Requirements of Target Solution}
\vspace{-0.5ex}
We derive \emph{four} specific requirements from the challenges listed above and expert feedback:
\textbf{Accuracy (R1):} answer accuracy $\geq 0.7$ required to be valuable to the use case.
\textbf{Transparency (R2):} link answers to source documents to support validation of response and increase users' trust.
\textbf{Continuous Updates (R3):} Knowledge must be easily updateable with new incidents such that the answers reflect the current state. 
\textbf{Scalability (R4):} The answer should come fast over a large corpus of data. The solution should be efficiently deployable.

\section{\emph{SuSi-RAG} for Intelligent Equipment Maintenance}\label{sec_solution}
\begin{figure*}[t]
\centerline{\includegraphics[width=0.92\textwidth]{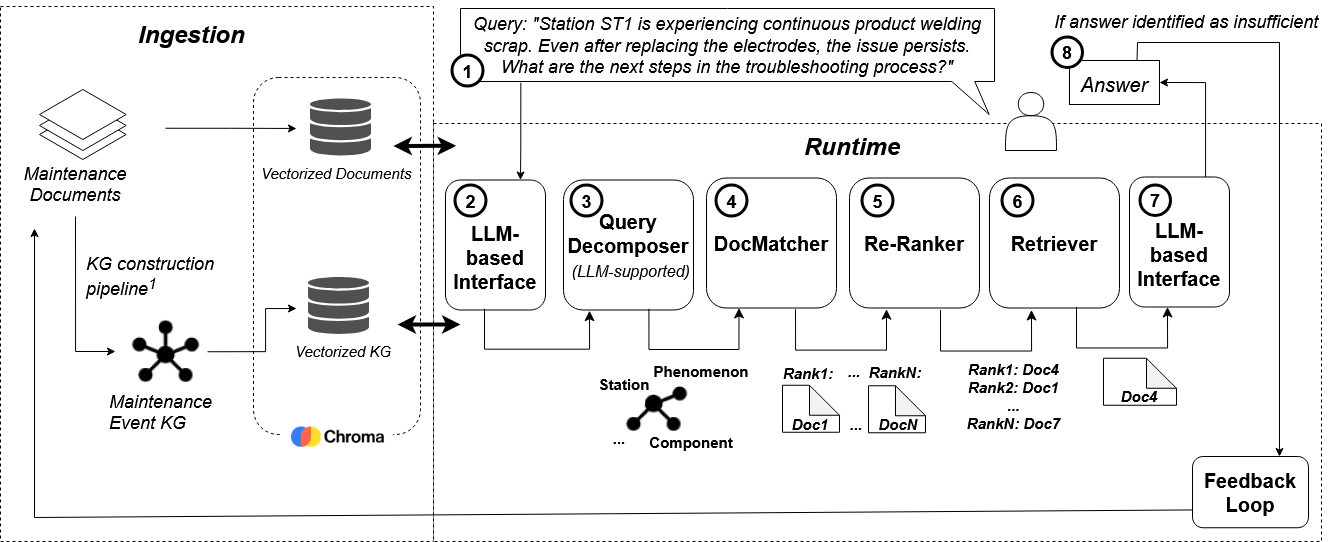}}
\caption{The ingestion and runtime overview of \emph{SuSi-RAG}.}
\label{overview}
\end{figure*}

Next, we explain why a hybrid architecture is needed, then describe the \emph{KG}, refer to the ingestion pipeline, and further introduce all runtime modules, see Figure~\ref{overview}. 
\vspace{-1ex}
\subsection{Hybrid Graph RAG Motivation}
LLMs provide a user-friendly, natural language interface, however, lack proprietary (maintenance) knowledge. 
RAG advantages~\cite{gao2023retrieval} are to improve accuracy (\textbf{R1}) and transparency (\textbf{R2}), maintain timeliness (\textbf{R3}), and improve scalability (\textbf{R4}). 
However, pure vector-based RAG solutions struggle with limited contextual awareness, noise in the retrieval process, and conflicting or non-factual information~(\cite{peng2025graph,liu2024lost}).
Graph RAG takes into account the relationships and interconnections within the data, enabling more accurate and comprehensive information retrieval especially for domain-specific scenarios.
While Graph RAG is strong on complex causal chains~(\textbf{C1}), it shows limitations on recent, unstructured information (e.g., latest maintenance incidents which are not yet in a KG) or detailed text instructions. Hence, we adopt a hybrid RAG (see e.g.,~\cite{sarmah2024hybridrag,10.1007/978-981-97-5678-0_34}), combining vector- and graph-based retrieval.

Considering Table~\ref{tab:typical_questions}, a graph RAG can identify in \textbf{C1} a \emph{cylinder failure} as a root cause and provide a diagnostic path to increase transparency to the user. An additional vector search on \emph{cylinder failure repair} can provide text-based repair instructions. While a graph RAG can identify and connect entities (e.g., \emph{mechanical symptom} with \emph{modifying process parameters via station analogy}) and retrieve \emph{similar stations} in \textbf{C2}, a vector-based RAG retrieves documents with instructions on these entities.
\emph{SuSi-RAG} integrates historical maintenance knowledge to support engineers’ daily maintenance tasks by providing precise suggestions for resolving machine errors on a selected Bosch production line. With a specific integration of KG modules \textbf{SuSi-RAG} enables effective similarity searches that incorporate semantic meaning and complex relationships, while reducing redundancy and noise in retrieved information.

\subsection{Ontology and Knowledge Graph}\label{subsec:ont_kg}
\begin{figure}[htbp]
\centerline{\includegraphics[width=\columnwidth]{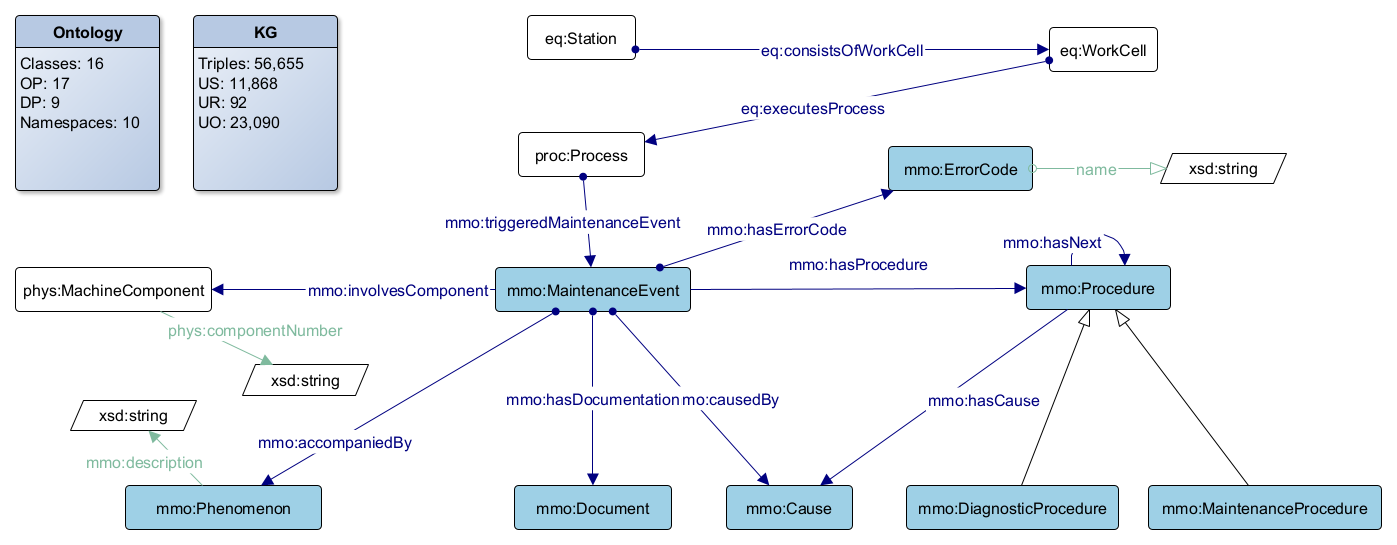}}
\caption{Main concepts of ontology \textbf{MMO}~\cite{zhang2025knowledge} and metrics of \textbf{MMO} and generated Maintenance KG. Key: OP: object properties; DP: datatype properties; US: unique subjects; UR: unique relationships; UO: unique objects.}
\label{fig:kg_metrics}
\end{figure}

Ontology development and semi-automated KG construction pipeline are shown in a previous work~\cite{zhang2025knowledge}. The ontology is a maintenance extension of an ontology set from the \emph{Bosch I4.0 Knowledge Graph}~\cite{grangel2020knowledge} which reuse most important concepts from the manufacturing domain, e.g., machines, components, plants, lines, among others\footnote{\url{https://github.com/eclipse-esmf/esmf-manufacturing-information-model}}. The ontology with its central class \texttt{MaintenanceEvent} which links, e.g., to classes \texttt{MachineComponent} and \texttt{ErrorCode} is shown in Figure~\ref{fig:kg_metrics} along with its metrics. 
\subsection{Ingestion Pipeline and Embedding}
The equipment maintenance documents are embedded directly into a vector-database, the populated KG is vectorized separately. Both databases are accessed during runtime by our solution.
Embedding-based retrieval is the foundation of our approach in document search. 
We use the \emph{Bilingual and Crosslingual Embedding for RAG (BCEmbedding)} model, a robust and efficient embedding framework specialized for RAG implementations with strong performance~\cite{youdao_bcembedding_2023}. This model is optimized for multilingual and domain-specific text, making it highly suitable for our use case. The approach encompasses an \emph{EmbeddingModel} and a \emph{RerankerModel}, working together to enhance retrieval precision and accuracy.

\vspace{-1ex}
\subsection{Runtime Modules}
\begin{figure}[htbp]
\centerline{\includegraphics[width=\columnwidth]{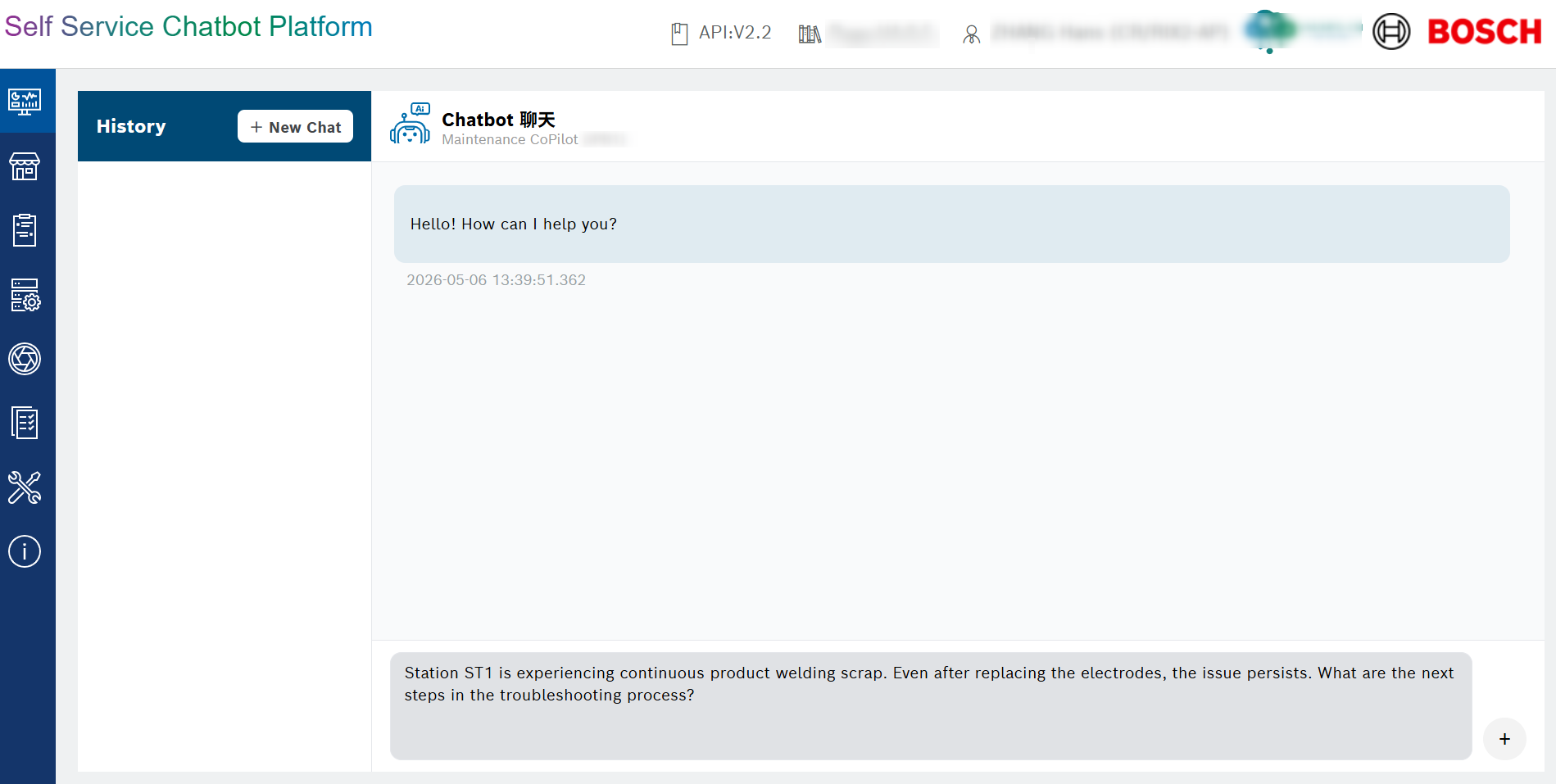}}
\caption{Screenshot of \emph{LLM-based Interface} of \emph{SuSi-RAG} during runtime.}
\label{fig:chatbot}
\vspace{-2ex}
\end{figure}

In the following part, we describe how a query is processed along the modules and introduce each module, see Figure~\ref{overview}.\\
\noindent
\textbf{LLM-based Interface.} First, a user, e.g., maintenance engineer, enters a question into the \emph{LLM-based Interface}. This natural language interface is a chatbot as shown in Figure~\ref{fig:chatbot}.
\noindent
\textbf{Query Decomposer.} 
The \emph{Query Decomposer} module receives the user question as input and first interprets, zero-shot, with an LLM agent whether the question relates to a maintenance issue (\emph{“True”} / \emph{“False”}), and, if true, extracts predefined maintenance-related key-value pairs. 
For the extraction process, the keys $k_i$, $i=1..n$, represent nodes associated with a maintenance event in the ontology, while the values $v_i$ are directly extracted from the question. If the key $k_i$ is not mentioned in the question, its corresponding value $v_i$ is “None”. For example, given a question $Q$ \emph{=“Station S1 alarmed, component C2: axis stopped. Crash. How to deal with it?”}, we extract keys including \emph{station}, \emph{phenomenon}, \emph{process}, \emph{machine component}, and \emph{error code}, with the corresponding results as follows: \emph{"station":"Station S1"}, \emph{"phenomenon description": "axes stopped. crash."}, \emph{"process": None}, \emph{"machine component":"component C2"}, \emph{"error code": None}. More keys (nodes) of our ontology are shown in Figure~\ref{fig:kg_metrics}. As illustrated in Figure~\ref{overview}, the extracted key-value pairs can be represented as a maintenance event subgraph, which the following modules use for comparison and search. 
For queries highly similar to previous ones, we reuse historical query decomposer results stored in the vector database and bypass the query decomposition process, significantly accelerating the pre-retrieval module.

\noindent
\textbf{Doc-Matcher.} 
Figure~\ref{overview} illustrates that the input of the \emph{Doc-Matcher} module is a parsed question $Q$ represented as a maintenance subgraph and further, each document $E_i$ is represented as a maintenance event subgraph. These subgraphs include nodes for maintenance events, stations, error codes, incident causes, and more. The important element are the node values. To identify the most relevant document, we develop a subgraph similarity score $S_{E_i}$ which enables fine-grained similarity matching for maintenance events. Specifically, the \emph{BCEmbedding} model is employed to generate vectorized representations of the node values. First, we calculate the matching score between each key in the decomposed query and its corresponding entity in each maintenance event subgraph. For textual entities such as \emph{phenomenon}, we use cosine similarity. For pattern-based entities, e.g., \emph{machine component}, which follow specific patterns of digits and letters, we apply regular expressions with a binary score: \emph{matched (1)} or \emph{not matched (0)}. To account for the varying importance of different entity types, a dynamic weighting factor is assigned to each entity type based on expert recommendations. 
As some maintenance documents may be missing or show empty values for certain entities, leading to discrepancies in the final matching scores, we normalize the weights by scaling each entity’s weight relative to the total weight of all extracted document entities. This ensures consistent and comparable matching scores, regardless of variations in the number of available entities. 

Each user query $Q$ can be decomposed into a set of key-value pairs $(k_j, v_j)$. Then, we define $w_{k_j}$ to be the weight of entity $k_j$ according to the experts' weight assignment of entity types. The total matching score $S_{E_i}$ of a user query $Q$ with a maintenance event $E_i$ is thus calculated as follows:
\begin{equation}\label{eq:score_raw}
\begin{footnotesize}
\begin{aligned}
S_{E_i} = \!\!\sum_{(k_j, v_j) \in Q} \biggl[ &
\left( I(\text{type}(k_j)=\text{pattern}) \cdot I(E_i.\text{sec} [ \text{value} ] = v_j) \cdot \text{norm}(E_i,w_{k_j}) \right. \\ & + 
\left. I(\text{type}(k_j)=\text{text}) \cdot \text{sim}(E_i.\text{sec} [ \text{value} ], v_j) \cdot \text{norm}(E_i,w_{k_j}) \right)
\biggr]
\end{aligned}
\end{footnotesize}
\end{equation}
with $E_i.sec$ the entity representing a key element in $E_i$, $E_i.sec[\text{value}]$ the value of entity $E_i.sec$, $\text{norm}(E_i,w_{k_j})$ the normalized weight for entity $k_j$ in $E_i$ given as follows: 
 $$\text{norm}(E_i,w_{k_j}) := \frac{w_{k_j}}{\sum_{k_l \in E_i, k_l \neq \emph{None}} w_{k_l}},$$
where $I(\text{condition})$ is a boolean indicator function which is $1$ if the condition inside the brackets is true, and $0$ otherwise, and $\text{sim}(E_i.sec [ \text{text} ], v_j)$: cosine similarity\footnote{See e.g., \url{https://en.wikipedia.org/wiki/Cosine_similarity}} for textual nodes.

To prevent a document with only a single matched entity to achieve a disproportionately high matching score, we apply a penalty factor to scale the raw matching score $S_{E_i}$ by the fraction of matched entity types. We denote this penalty factor with \emph{PF} and compute the final scaled score $S^{'}_{E_i}$ as:
\begin{equation}
    S^{'}_{E_i} = S_{E_i} \cdot \frac{1}{1 + (\frac{1}{|matched \_ sec(E_i)|} - \frac{1}{|all \_ sec|})}, 
\end{equation}
where $S_{E_i}$ is the raw matching score of maintenance event $E_i$ as calculated in~(\ref{eq:score_raw}), $|matched \_ sec(E_i)|$ is the number of matched entities in maintenance event $E_i$ (i.e., values for the entities exist in query $Q$), and $|all \_ sec|$ is the number of all entities in $E_i$.
\begin{algorithm}[t!]
    \caption{Document Matcher}\label{alg:docmatcher}
    \begin{algorithmic}[1]
        \Require \textbf{Input:} Decomposed query $Q$, maintenance event set $E$
        \Ensure \textbf{Output:} Document matching score list $S_E$
        \State Initialize the score $S_{E_i} \xleftarrow{} 0$ for each maintenance event $E_i \in E$
        \For{each $E_i$ in $E$}
            \For{each $(k_j,v_j)$ in $Q$}
                \If{$\texttt{type}(k_j)=\texttt{pattern}$}
                    \State Check if $E_i.sec = k_j$
                    \State $s \xleftarrow{} I(E_i.sec[\texttt{value}]=v_j)$ \{Exact match for pattern-based entities\}
                \ElsIf{$\texttt{type}(k_j)=\texttt{textual}$} 
                    \State Check if $E_i.sec = k_j$
                    \State $s \xleftarrow{} \texttt{sim}(\texttt{embed}(E_i.sec[\texttt{text}]), \texttt{embed}(v_j))$
                \EndIf 
                \State $s \xleftarrow{} s \times \text{norm}(E_i,w_{k_j})$
                \State $S_{E_i} \xleftarrow{} S_{E_i} + s$
            \EndFor
            \State $S^{'}_{E_i} \xleftarrow{} S_{E_i} \times PF$
            \State Append $S^{'}_{E_i}$ to $S_E$
        \EndFor{}
        \State \Return Document matching score list $S_E$
    \end{algorithmic}
\end{algorithm}

To compute the desired output efficiently, we follow the procedure described in Algorithm~\ref{alg:docmatcher} and further store vectorized KG nodes and values in Chroma\footnote{https://docs.trychroma.com/} to speed up retrieval.

\noindent
\textbf{Re-Ranker and Retriever.} 
The \emph{Re-Ranker} module refines the results by re-scoring the top $k$ documents. For this, we recalculate the similarity between the user query and the documents with each the \emph{BM25}\footnote{\url{https://pypi.org/project/BM25/}} matching method and the \emph{BCE-RerankerModel}. The \emph{BCE-RerankerModel} jointly encodes the query and each document as a single sequence, resulting in richer token-level interactions than a dual encoder.
With the \emph{Reciprocal Rank Fusion (RRF)} method, we integrate the similarity scores from the subgraph-based, \emph{BM25} (with a slightly higher weight than the others), and \emph{BCE RerankerModel}, to combine rankings from all three methods into one robust final ranking.
The \emph{Retriever} fetches the corresponding subgraph (i.e., maintenance event) from the KG according to the \emph{Re-Ranker} results. A subgraph with higher relevance scores is considered more closely related to the user’s query. Using \emph{SPARQL} query templates, we extract all relevant information about the selected maintenance event as \emph{RDF} data from the KG. 
A post-processing step is applied to compress, simplify, and logically order the \emph{RDF} data, which often contains redundant information or verbose \emph{URIs}. The post-processed data is concise and structured for efficient LLM consumption. \\
\noindent
\textbf{Feedback Loop.} 
An optional \emph{feedback loop} provides users with the possibility to flag insufficient answers in two ways: as \emph{Knowledge Gaps} or \emph{Error Mapping Issues} of the system. Both flags trigger manual tasks, as described in Section~\ref{sec:eval}.
\section{Evaluation}\label{sec:eval}
\vspace{-0.5ex}
We conduct a two-fold evaluation: a controlled quantitative \emph{PoC} analysis to establish a baseline performance, and a continuous analysis on a Bosch production line for \emph{eight} months.

\subsection{Reproducibility and Setup}
Table~\ref{tab:settings} shows the \emph{PoC} configuration with LLM \emph{GPT 4.5}, embedding and reranking via \emph{BCEmbedding}, vector store Chroma and SPARQL queries to the KG.  

\begin{table}[htbp]
\caption{SuSi-RAG \emph{PoC} Implementation configuration.}
\begin{center}
\renewcommand{\arraystretch}{1.15}
\begin{tabular}{l p{0.45\columnwidth}}
\toprule
\textbf{Component}&\textbf{Configuration} \\
\midrule
Query decomposition \& intention & \emph{GPT 4.5} \\
Generation LLM & \emph{GPT 4.5} \\
LLM temperature & \emph{0.1} \\
Embedding& \emph{BCEmbedding} \\
Vector Store& \emph{Chroma} \\
Document Chunk Size& \emph{500 characters} \\
Document Chunk Overlap& \emph{100 characters} \\
\makecell[l]{Entity Weights (by Experts)\\ \\ \\ \\ } & \makecell[l]{\emph{phenomenon\_description}: 0.40, \\
                            \emph{component}: 0.15, \emph{errorcode}: 0.15,  \\
                           \emph{station}: 0.10, \emph{process}: 0.10,  \\
                           \emph{line}: 0.10 }  \\
\makecell[l]{RRF Constant of Source \\ \\ } & \makecell[l]{Subgraph: 45, BM25 ReRank: 60, \\BCE ReRanker: 40} \\
\bottomrule
\end{tabular}
\label{tab:settings}
\end{center}
\vspace{-2ex}
\end{table}

\subsection{PoC Evaluation}
We evaluate the \emph{PoC} based on 26 expert-curated maintenance queries which originate via expert-verified source documents and reference answers from six production stations and diverse fault types, including sensor, communication, motion, mechanical, screw-handling and software incidents. 
The retrieval performance measures the system's ability to identify the correct source document within the top-\emph{k} results. Table~\ref{tab:retrieval_eval} shows that \emph{SuSi-RAG} with 96.2\% (25/26~queries; 95\%~Wilson, Confidence Interval~(CI):~81.1\%-99.3\%) outperforms in Top-1 a vector-based RAG baseline including reranking and implemented via \emph{Azure AI Search} with 61.5\% (16/26~queries; 95\%~Wilson, CI:~42.5\%-77.6\%). \emph{SuSi-RAG's} high top-1 score is critical to reduce the cognitive load on shopfloor personnel during time-sensitive maintenance tasks. On 26 queries \emph{SuSi-RAG} answers in nine more cases correctly in which the baseline RAG fails, while no new failures are introduced. A partial ablation, i.e., removing only the reranking component (\emph{SuSi-RAG} excl. reranking, 18/26; 95\%~Wilson, CI: 50.0\%-83.5\%), shows a value add of the graph-based retrieval over the baseline (incl. reranking) as well as a clear value add of the reranking component in \emph{SuSi-RAG}. Note that the CIs of the baseline and of \emph{SuSi-RAG} excl. reranking overlap, so this difference is indicative only with 26 queries. This hints at complementary effects of structured retrieval, e.g., via graph-based retrieval and expert weights, to improve the candidate set, and cross-encoder reranking to push the correct document to the top 1 position. Other components (expert weights, penalty factor, BM25) are not ablated individually. A full ablation study is future work.

We evaluated the generated answers on expert correctness score (answer quality), completeness, and cosine similarity to the expert answers. The expert correctness score is manually rated with 0 (incorrect), 1, 2, or 3 (fully correct) by a domain expert against the gold answer. The reported percentage is the sum of ratings divided by the maximum score across 23 queries ($69=3*23$). Three queries from the 26 had to be excluded as for one query the baseline answer was not evaluated by the experts by mistake, with no possibility to rerun the experiment with the same experts, and two queries were added to the set after the expert evaluation, again with no possibility to rerun with the same experts. The expert correctness score is evaluated on the same 23 queries for both systems, all other results were run based on 26 queries. Completeness is calculated as the proportion of elements in the generated answer that match expert-identified key elements in the gold standard answer. Cosine similarity measures the similarity between the embedding representations of the generated and gold standard answers.  
The expert correctness score with \emph{SuSi-RAG} with 84.1\% nearly doubles compared to the baseline RAG as shown in Table~\ref{tab:generation_eval}. Cosine similarity and, especially, the completeness are increased with \emph{SuSi-RAG}.

\begin{table}[ht]
\centering
\caption{Evaluation of retrieval process (in \%). Key: n = number of evaluated queries.}
\label{tab:retrieval_eval}
\begin{tabular}{lcccc}
\toprule
\textbf{Configuration} & \multicolumn{2}{c}{\textbf{Top-1} (n=26)} & \multicolumn{2}{c}{\textbf{Top-3} (n=26)} \\ 
 & Queries & \% & Queries & \% \\ 
\midrule
Baseline RAG & 16/26 & 61.5\% & 24/26 & 92.3\% \\
SuSi-RAG excl. reranking & 18/26 & 69.2\% & 25/26 & 96.2\% \\
\textbf{SuSi-RAG} & \textbf{25/26} & \textbf{96.2\%} & \textbf{26/26} & \textbf{100\%} \\ 
\bottomrule
\end{tabular}
\vspace{-2ex}
\end{table}
\begin{table}[ht]
\centering
\caption{Evaluation of answer generation (in \%). Key: n = number of evaluated queries.}\label{tab:generation_eval}
\newcolumntype{Y}{>{\centering\arraybackslash}X}
\begin{tabularx}{\linewidth}{l Y Y Y}
\toprule
\textbf{Approach} & \textbf{Expert Correctness Score} (n=23) & \textbf{Completeness} (n=26) & \textbf{Cosine Sim.} (n=26) \\ \midrule
Baseline RAG & 44.9\% & 38.0\% & 73.2\% \\
\textbf{SuSi-RAG} & \textbf{84.1\%} & \textbf{79.6\%} & \textbf{83.2\%} \\
\bottomrule
\end{tabularx}
\vspace{-2ex}
\end{table}
\subsection{In-Use Production Analysis}
\begin{figure}[ht]
\centerline{\includegraphics[width=\linewidth]{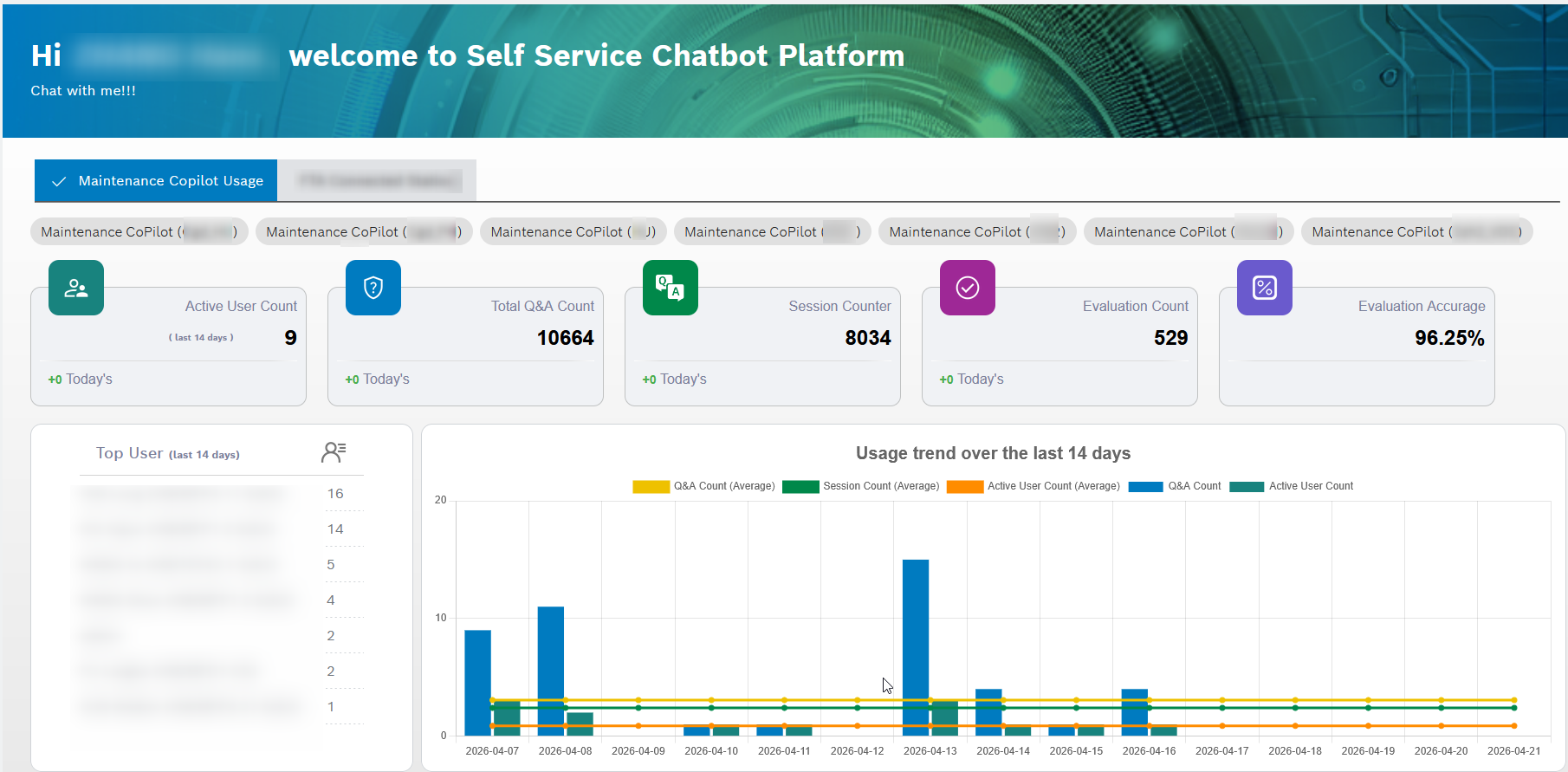}}
\caption{The Maintenance Chatbot dashboard developed by plant personnel for tracking real-world usage and accuracy metrics.}
\label{fig:dashboard}
\end{figure}

\begin{figure}[ht]
\centerline{\includegraphics[width=\linewidth]{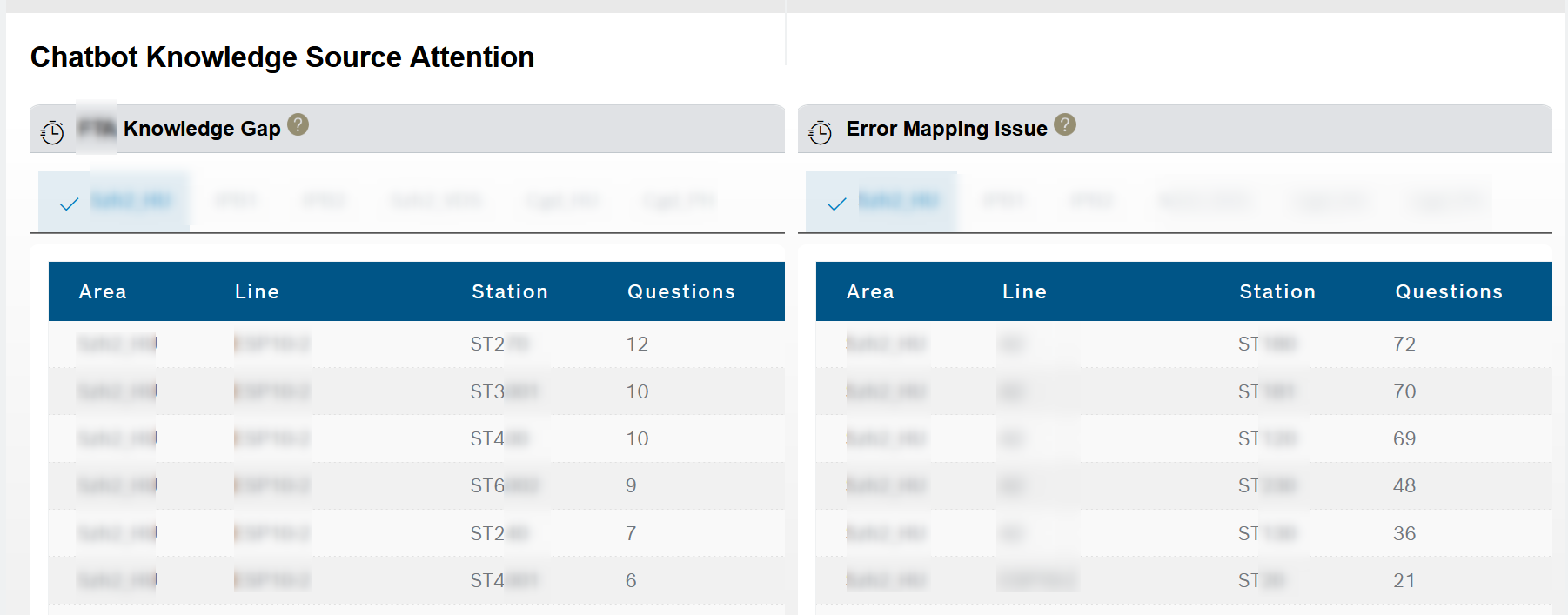}}
\caption{Feedback monitor to identify and bridge maintenance knowledge gaps.}
\label{fig:knowledge_GAP}
\vspace{-3ex}
\end{figure}

\emph{SuSi-RAG} has been running on an active Bosch production line for \emph{eight} months. After initial deployment, the core development team handed \emph{SuSi-RAG} over to the operational IT plant team, which developed a specialized \emph{Maintenance Chatbot} dashboard (Figure~\ref{fig:dashboard}) to track key performance indicators and generate operational insights.
The system's performance in the production environment shows the following characteristics:
\begin{itemize}
    \item \textbf{Usage Volume:} As of April 2026, the system has successfully handled \emph{10,664} Q\&A interactions across \emph{8,034} individual sessions.
    \item \textbf{Confirmed Error Rate:} Of 10,664 answers, users flagged 529 for review; maintenance experts confirmed 400 of them as incorrect, yielding a confirmed error rate of 3.75\% (95\% Wilson, CI: 3.41-4.13\%). Since unflagged answers were not verified, 96.25\% is an upper bound on accuracy rather than a verified value, yet reflecting the actual perceived reliability of the system by users.
\end{itemize}
The feedback loop ensures continuous evolvement of the KG. We monitor two primary metrics (see Figure~\ref{fig:knowledge_GAP}): 1. \textbf{Knowledge Gaps} identify specific Area-Line-Station combinations where historical data has not been processed and, hence, not fed into the KG. This alerts the IT plant team to prioritize data ingestion for those stations. 2. \textbf{Error Mapping Issues} occur when users ask questions for which no relevant historical maintenance issue can be identified in the KG. This signals a need to either correct existing triples or ingest new "gold answers" from recent incidents.
\vspace{-0.5ex}

\subsection{Limitations}\label{subsec:limitations}
Despite usefulness in practice, our work shows the following limitations: \\
\noindent
\textbf{Ablation.} The baseline comparison against a vector-based approach and \emph{SuSi-RAG} without reranking limits the value add analysis of isolated components. A controlled, enhanced study with ablations is the most important next step.\\
\noindent
\textbf{Sample Size.} The verified PoC set of \emph{26} queries is small, yet, it is challenging, due to time and cost pressure, to collect more expert-labelled data.\\
\noindent
\textbf{Expert Verification of Flagged Answers.} The 529 answers in-use are flagged by users. Unflagged answers were not verified, hence, further errors may exist. \\
\noindent
\textbf{Reproducibility.} While we show configuration settings and parameters, see Table~\ref{tab:settings}, we may not disclose actual data and code of \emph{SuSi-RAG} due to proprietary reasons.
\section{Lessons Learned}\label{sec_lessonslearned}
We discuss five lessons learned and add our deployment experience of \emph{SuSi-RAG} at Bosch, spanning over \emph{eight} months and exceeding \emph{10,000} interactions, yielding critical insights into semantic maintenance systems lifecycles in Industry 4.0. 

\subsection{Granularity and Precision through Subgraph Matching}
A primary driver of our system's success is the transition from coarse-grained text chunk matching to fine-grained maintenance event subgraph matching. By incorporating domain-specific node weights and penalty factors for partial matches, we achieve a near-perfect retrieval accuracy in the \emph{PoC}. The quantitative results confirm that for complex equipment maintenance, the \emph{structural context} provided by a KG is more effective than pure similarity in vector space. Some key procedures are embedded within images which we address in future work (see Section~\ref{sec_conclusion}).

\vspace{-1ex}
\subsection{The Shift from Static Knowledge to an Evolving Ecosystem}
The eight-month deployment underlines the need of continuous updates for industrial KGs. Instead of reactive documentation, we utilize the \emph{Maintenance Chatbot} dashboard to drive a \emph{Continuous Knowledge Lifecycle}:\\
\noindent
\textbf{Usage-Driven Formalization}: By monitoring \emph{KGs}, the user demand now dictates the priority of formalization. This transforms the KG from a passive tool into a strategic driver, focusing expert effort on the most frequent information voids. \\
\noindent
\textbf{Managing Semantic Debt}: We use \emph{Error Mapping Issues} to quantify \emph{semantic debt}, the discrepancy between codified triples and shopfloor incidents. Logging these as \emph{gold answers} triggers a self-correcting cycle, ensuring synchronized KG evolution with machinery updates.\\
This proactive loop ensures the system remains resilient in a high-volume production environment, sustaining an upper bound answer accuracy of \emph{96.25\%} across \emph{10,664} interactions.

\vspace{-1ex}
\subsection{Speed and Volume Adjustments (R4)}
\vspace{-1ex}
This part summarizes \emph{three} minor improvements to meet speed and volume requirements: (i) to achieve a small time to answer~(\textbf{R4}), we significantly increase inference speed during query decomposition by reusing historical query decomposer results stored in the vector database (see \emph{Query Decomposer} in Section~\ref{sec_solution}); (ii) as similarity search can become slow, we increase speed by the KG vectorization (see \emph{Doc-Matcher} in Section~\ref{sec_solution}); (iii) we compress \emph{RDF} data before prompting to the LLM (see \emph{Re-Ranker and Retriever} in Section~\ref{sec_solution}).

\vspace{-1ex}
\subsection{Requirements Fulfillment}
\vspace{-1ex}
\textbf{R1} is met in the \emph{PoC} where the expert-rated answer quality (84.1\%) exceeds the required 0.7. After evaluating the in-use deployment of \emph{eight} months, we observe that the upper bound accuracy of 96.25\% and a confirmed error rate of 3.75\% indicate that \textbf{R1} is met. A verified accuracy requires a follow-up study (see Section~\ref{subsec:limitations}). The source links provided in the answers satisfy \textbf{R2}. Especially the feedback loop fulfills \textbf{R3}. Finally, the usage-driven formalization and speed and volume adjustments satisfy \textbf{R4} sufficiently.

\section{Conclusion and Future Steps}\label{sec_conclusion}
We introduce \emph{SuSi-RAG}, a novel hybrid graph RAG solution, to address critical knowledge access and utilization challenges within equipment maintenance at Bosch. In its core, \emph{SuSi-RAG} contains a maintenance event subgraph similarity matching algorithm with expert-weighted entities and a partial-match penalty factor, complemented by vector retrieval. \emph{SuSi-RAG} effectively handles issues arising from complex causal chains and cross-domain information resulting in a nearly doubled expert-rated answer quality from 44.9\% to 84.1\% and an increased retrieval accuracy from 61.5\% to 96.2\% in a \emph{PoC} setting.
We show practical value in a successful \emph{eight} month deployment on a Bosch production line where \emph{SuSi-RAG} handled more than \emph{10,000} interactions, with an upper bound answer accuracy of \emph{96.25\%} (confirmed error rate 3.75\%). 
The modular design of \emph{SuSi-RAG} comprises an \emph{LLM-based Interface}, \emph{four} main sequential query-handling modules, and a fifth module for continuous data quality improvement. \emph{SuSi-RAG} ensures high-precision responses and offers advantage of reusable components for other RAG systems within Bosch.
In the future, we target the systems' limitations. First, we increase the number of expert-verified maintenance queries to increase statistical reliability of the performance. Second, we conduct a controlled full ablation study to identify the value-add components. Third, we include multi-modal data, starting with images which are helpful in complex maintenance tasks. Fourth, we explore agent integration, e.g., for complex reasoning tasks to develop suggestions on KG completion. These suggestions can after expert review be added to the KG without a prior, actual incident. Fifth, we enhance explainability by adding retrieval path and justification for the ranking.

\section*{Acknowledgements}
\vspace{-0.5ex}
The authors thank Qiushi Cao, Tianyang Liu, Sebastian Monka and Irl{\'{a}}n Grangel{-}Gonz{\'{a}}lez for valuable exchanges on this work. 

\vspace{-1ex}
\section*{Declaration on Generative AI}
\vspace{-0.5ex}
During the preparation of this work, the authors used Gemini Enterprise~2.5~Pro (with Nano Banana 2.5) for Figure~\ref{fig:motivation} in order to: Generate large portions of the image. The authors further used Gemini Enterprise~2.5~Pro on parts of the paper in order to: Rephrase small portions of the text. After using this tool, the authors reviewed and edited the content as needed and take full responsibility for the publication’s content. 

\bibliographystyle{IEEEtran}
\bibliography{IEEEabrv,bibliography}

\begin{thebibliography}{10}
\providecommand{\url}[1]{#1}
\csname url@samestyle\endcsname
\providecommand{\newblock}{\relax}
\providecommand{\bibinfo}[2]{#2}
\providecommand{\BIBentrySTDinterwordspacing}{\spaceskip=0pt\relax}
\providecommand{\BIBentryALTinterwordstretchfactor}{4}
\providecommand{\BIBentryALTinterwordspacing}{\spaceskip=\fontdimen2\font plus
\BIBentryALTinterwordstretchfactor\fontdimen3\font minus \fontdimen4\font\relax}
\providecommand{\BIBforeignlanguage}[2]{{%
\expandafter\ifx\csname l@#1\endcsname\relax
\typeout{** WARNING: IEEEtran.bst: No hyphenation pattern has been}%
\typeout{** loaded for the language `#1'. Using the pattern for}%
\typeout{** the default language instead.}%
\else
\language=\csname l@#1\endcsname
\fi
#2}}
\providecommand{\BIBdecl}{\relax}
\BIBdecl

\bibitem{lasi2014industry}
H.~Lasi, P.~Fettke, H.-G. Kemper, T.~Feld, and M.~Hoffmann, ``Industry 4.0,'' \emph{Business \& information systems engineering}, vol.~6, no.~4, 2014.

\bibitem{lin2023industry}
B.~Lin, W.~Wu, and M.~Song, ``Industry 4.0: Driving factors and impacts on firm’s performance: An empirical study on china’s manufacturing industry,'' \emph{Annals of Operations Research}, vol. 329, no.~1, 2023.

\bibitem{DBLP:journals/cii/SilvestriFISC20}
L.~Silvestri, A.~Forcina, V.~Introna, A.~Santolamazza, and V.~Cesarotti, ``Maintenance transformation through industry 4.0 technologies: {A} systematic literature review,'' \emph{Computers in Industry}, vol. 123, 2020.

\bibitem{iglesias2022ontology}
A.~Iglesias~Molina, J.~A. Bernab{\'e}~Diaz, P.~Deshmukh, P.~P. Espinoza~Arias, A.~Ayll{\'o}n-Ben{\'\i}tez, A.~Fern{\'a}ndez~Izquierdo, J.~M. Ponce-Bernab{\'e}, S.~P{\'e}rez, E.~Ruckhaus, O.~Corcho \emph{et~al.}, ``Ontology management in an industrial environment: The basf governance operational model for ontologies (gomo),'' \emph{2022 ISMB Bio-Ontologies}, 2022.

\bibitem{yahya2024benchmark}
M.~Yahya, A.~Ali, Q.~Mehmood, L.~Yang, J.~G. Breslin, and M.~I. Ali, ``A benchmark dataset with knowledge graph generation for industry 4.0 production lines,'' \emph{Semantic Web}, vol.~15, no.~2, 2024.

\bibitem{hofer2024construction}
M.~Hofer, D.~Obraczka, A.~Saeedi, H.~K{\"o}pcke, and E.~Rahm, ``Construction of knowledge graphs: Current state and challenges,'' \emph{Information}, vol.~15, no.~8, 2024.

\bibitem{kalayci2020semantic}
E.~G. Kalayc{\i}, I.~Grangel~Gonz{\'a}lez, F.~L{\"o}sch, G.~Xiao, A.~ul~Mehdi, E.~Kharlamov, and D.~Calvanese, ``Semantic integration of bosch manufacturing data using virtual knowledge graphs,'' in \emph{International Semantic Web Conference}.\hskip 1em plus 0.5em minus 0.4em\relax Springer, 2020.

\bibitem{DBLP:conf/eurocast/ChhetriAFGGG22}
\BIBentryALTinterwordspacing
T.~R. Chhetri, S.~Aghaei, A.~Fensel, U.~G{\"{o}}hner, S.~G{\"{u}}l{-}Ficici, and J.~Martinez{-}Gil, ``Optimising manufacturing process with bayesian structure learning and knowledge graphs,'' in \emph{Computer Aided Systems Theory - {EUROCAST} 2022 - 18th International Conference, Revised Selected Papers}, vol. 13789.\hskip 1em plus 0.5em minus 0.4em\relax Springer, 2022. [Online]. Available: \url{https://doi.org/10.1007/978-3-031-25312-6\_70}
\BIBentrySTDinterwordspacing

\bibitem{hu2025grag}
Y.~Hu, Z.~Lei, Z.~Zhang, B.~Pan, C.~Ling, and L.~Zhao, ``Grag: Graph retrieval-augmented generation,'' in \emph{Findings of the Association for Computational Linguistics: NAACL 2025}, 2025.

\bibitem{xu2024retrieval}
Z.~Xu, M.~J. Cruz, M.~Guevara, T.~Wang, M.~Deshpande, X.~Wang, and Z.~Li, ``Retrieval-augmented generation with knowledge graphs for customer service question answering,'' in \emph{Proceedings of the 47th international ACM SIGIR conference on research and development in information retrieval}, 2024.

\bibitem{gao2023retrieval}
Y.~Gao, Y.~Xiong, X.~Gao, K.~Jia, J.~Pan, Y.~Bi, J.~Sun, H.~Wang \emph{et~al.}, ``Retrieval-augmented generation for large language models: A survey,'' \emph{arXiv preprint arXiv:2312.10997}, 2023.

\bibitem{lewis2020retrieval}
P.~Lewis, E.~Perez, A.~Piktus, F.~Petroni, V.~Karpukhin, N.~Goyal, H.~K{\"u}ttler, M.~Lewis, W.-t. Yih, T.~Rockt{\"a}schel \emph{et~al.}, ``Retrieval-augmented generation for knowledge-intensive nlp tasks,'' \emph{Advances in neural information processing systems}, vol.~33, 2020.

\bibitem{peng2025graph}
B.~Peng, Y.~Zhu, Y.~Liu, X.~Bo, H.~Shi, C.~Hong, Y.~Zhang, and S.~Tang, ``Graph retrieval-augmented generation: A survey,'' \emph{ACM Transactions on Information Systems}, vol.~44, no.~2, 2025.

\bibitem{edge2024local}
D.~Edge, H.~Trinh, N.~Cheng, J.~Bradley, A.~Chao, A.~Mody, S.~Truitt, D.~Metropolitansky, R.~O. Ness, and J.~Larson, ``From local to global: A graph rag approach to query-focused summarization,'' \emph{arXiv preprint arXiv:2404.16130}, 2024.

\bibitem{li2024simple}
M.~Li, S.~Miao, and P.~Li, ``Simple is effective: The roles of graphs and large language models in knowledge-graph-based retrieval-augmented generation,'' \emph{arXiv preprint arXiv:2410.20724}, 2024.

\bibitem{sarmah2024hybridrag}
B.~Sarmah, D.~Mehta, B.~Hall, R.~Rao, S.~Patel, and S.~Pasquali, ``Hybridrag: Integrating knowledge graphs and vector retrieval augmented generation for efficient information extraction,'' in \emph{Proceedings of the 5th ACM International Conference on AI in Finance}, 2024.

\bibitem{10.1007/978-981-97-5678-0_34}
S.~Xu, M.~Chen, and S.~Chen, ``Enhancing retrieval-augmented generation models with knowledge graphs: Innovative practices through a dual-pathway approach,'' in \emph{Advanced Intelligent Computing Technology and Applications}.\hskip 1em plus 0.5em minus 0.4em\relax Springer Nature Singapore, 2024.

\bibitem{zhou2026lightweight}
T.~Zhou, Y.~Liu, M.~Kumar, and L.~Zou, ``Lightweight query-adaptive rag framework for knowledge support in smart manufacturing,'' \emph{Journal of Manufacturing Systems}, vol.~86, 2026.

\bibitem{CHANDRASEKHAR2024100232}
\BIBentryALTinterwordspacing
A.~Chandrasekhar, J.~Chan, F.~Ogoke, O.~Ajenifujah, and A.~{Barati Farimani}, ``Amgpt: A large language model for contextual querying in additive manufacturing,'' \emph{Additive Manufacturing Letters}, vol.~11, 2024. [Online]. Available: \url{https://www.sciencedirect.com/science/article/pii/S2772369024000409}
\BIBentrySTDinterwordspacing

\bibitem{manzini2010maintenance}
R.~Manzini, A.~Regattieri, H.~Pham, and E.~Ferrari, \emph{Maintenance for industrial systems}.\hskip 1em plus 0.5em minus 0.4em\relax Springer, 2010.

\bibitem{zhang2025knowledge}
\BIBentryALTinterwordspacing
H.~Zhang, W.~J. Schmidt, X.~Shen, Q.~Cao, S.~Monka, and A.~Paschke, ``Knowledge graph construction towards a graph rag-enhanced intelligent maintenance chatbot,'' in \emph{International Workshop on Scaling Knowledge Graphs for Industry 2025}, 2025. [Online]. Available: \url{https://ceur-ws.org/Vol-4064/SKGi-paper3.pdf}
\BIBentrySTDinterwordspacing

\bibitem{wang2025llm}
F.~Wang, D.~Shi, J.~Aguilar, X.~Cui, J.~Jiang, L.~Shen, and M.~Li, ``Llm-kgmqa: large language model-augmented multi-hop question-answering system based on knowledge graph in medical field,'' \emph{Knowledge and Information Systems}, vol.~67, no.~8, 2025.

\bibitem{karim2024scalable}
M.~R. Karim, F.~Hermsen, S.~A. Chala, P.~De~Perthuis, and A.~Mandal, ``Scalable semi-supervised graph learning techniques for anti money laundering,'' \emph{IEEE Access}, vol.~12, 2024.

\bibitem{liu2024lost}
N.~F. Liu, K.~Lin, J.~Hewitt, A.~Paranjape, M.~Bevilacqua, F.~Petroni, and P.~Liang, ``Lost in the middle: How language models use long contexts,'' \emph{Transactions of the association for computational linguistics}, 2024.

\bibitem{grangel2020knowledge}
I.~Grangel-Gonz{\'a}lez, F.~L{\"o}sch, and A.~ul~Mehdi, ``Knowledge graphs for efficient integration and access of manufacturing data,'' in \emph{2020 25th IEEE International Conference on Emerging Technologies and Factory Automation (ETFA)}, vol.~1, 2020.

\bibitem{youdao_bcembedding_2023}
N.~Youdao, ``Bcembedding: Bilingual and crosslingual embedding for rag,'' \url{https://github.com/netease-youdao/BCEmbedding}, 2023.

\end{thebibliography}

\end{document}